\documentclass[12pt]{book}

\usepackage[dvips]{graphicx,color}
\usepackage{makeidx,tsukuba}

\makeauthorindex
\makeindex

\begin{document}

\BookTitle{\itshape The 28th International Cosmic Ray Conference}
\CopyRight{\copyright 2003 by Universal Academy Press, Inc.}
\pagenumbering{arabic}

\chapter{Control Software for the VERITAS \v{C}erenkov Telescope System}
\author{
%
%
H.~Krawczynski$^1$, M.~Olevitch$^1$, G.~Sembroski$^2$, K.~Gibbs$^3$\\
\it
(1) Washington University, St. Louis, MO, USA \\
(2) Purdue University, West Lafayette, IN, USA \\
(3) Smithsonian Astrophysical Observatory, USA 
}
\section*{Abstract}
The VERITAS collaboration is developing a system of initially 4 and 
eventually 7 \v{C}erenkov Telescopes of the 12 m diameter class for high
sensitivity gamma-ray astronomy in the $>$50 GeV energy range. 
In this contribution we describe the software that controls 
and monitors the various VERITAS sub-systems.
The software uses an object-oriented approach to cope with the 
complexities that arise from using sub-groups of the 7 VERITAS
telescopes to observe several sources at the same time.
Inter-process communication is based on the CORBA Object 
Request Broker protocol and watch-dog processes
monitor the sub-system performance.
\section{Introduction}
Systems of Imaging Atmospheric \v{C}erenkov Telescopes have emerged as the 
technique of choice to detect gamma rays in the energy region from
50 GeV to 50 TeV. The images of air showers taken with fast cameras made
of typically 500 Photo-Multiplier Tubes (PMTs) within a couple of 
nano-seconds allow one to determine the direction and energy of individual 
GeV photons to an accuracy of 6 arcmin and 10\%, respectively, and to achieve
a high 100 GeV sensitivity of $\simeq$5 milli-Crab (4 VERITAS telescopes,
50 hrs on-source). 

\begin{table}[t]
 \caption{List of VERITAS sub-systems.}
\begin{center}
{
\begin{tabular}{|p{4cm}|p{9.3cm}|}
\hline
Name  & Function  \\
\hline

Telescope Positioner       & Tracks celestial sources with sub-groups of telescopes.\\
Central Trigger            & Fires on coincident telescopes triggers.\\
Data Harvester             & Collects data from all telescopes and builds events.\\
Central Calibration        & Sends calibration laser pulses to telescopes.\\
Atmospheric Monitors       & Various systems to monitor the atmosphere.\\
Data Base/Archive          & Archive housekeeping, science and calibration data.\\
Telescope Data Acquisition & Collects data at individual telescopes.\\
VME Readout                & Reads out flash-digitized data.\\
Charge Injection           & Calibrates readout electronics and pattern trigger.\\
\hline
\end{tabular}
}
\end{center}
\end{table}
The operation of the experiments requires the following major software 
tasks: (i) control of the various hardware components, 
(ii) readout of the digitized data, (iii) archiving of data and 
housekeeping information, and (iv) data reduction.
In this paper we describe the implementation of the first of these tasks
for the VERITAS \v{C}erenkov telescope system.
A certain complexity of this task arises from the large number of
sub-systems that are involved in the operation of the telescopes
(Table 1).
The VERITAS telescope system is described by Wakely et al., in these proceedings.
Other aspects of the telescope software are given in the contributions by
Cui et al., Kosack et al., and Fegan and et al.
\section{The Central Control Software}
\begin{figure}[t]
  \begin{center}
    \includegraphics[height=24.5pc]{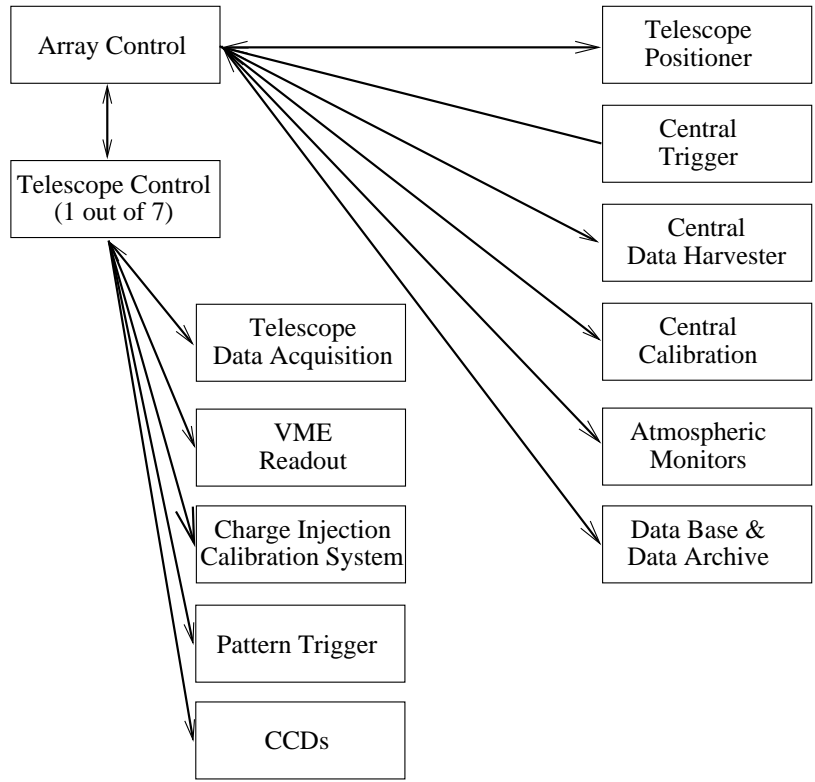}
  \end{center}
  \vspace{-0.5pc}
  \caption{Control hierarchy of VERITAS software processes.
The program Array Control launches and monitors all VERITAS sub-systems.
Processes that run at individual telescopes, are controlled by the program
``Telescope Control''.}
  \label{ac}
\end{figure}
The VERITAS software is hierarchically structured as shown in Fig.\ \ref{ac}.
The user controls the experiment via the program ``Array Control''. 
This program controls a number of ``global'' tasks that are unique to the 
experiment including the central trigger system, a central calibration
system, the data harvester program, the telescope positioner system, and
atmospheric monitoring systems.
Furthermore, it controls the so-called ``Telescope Control'' programs
that run on computers located at each telescope (telescopes
are located at a next-neighbor distance of $\sim$100 m).

The Telescope Control programs monitor 
all the systems that run locally at the individual telescopes, 
namely the local data acquisition, the PMT High Voltage and Current Monitor 
systems, the local trigger logic, a charge injection system used 
for calibration of the readout electronics, the telescope trigger logic, and
CCDs used to determine the pointing of the telescopes.
The telescope positioner system would naturally fit into the hierarchy 
``below'' the level of the Telescope Control program; however,
for security reasons, we decided to treat it as an independent 
``global'' system.

The software is designed to support the usage of arbitrary sub-telescope 
systems. For example 2 times 2 telescopes might be used to track 2 sources
in stereoscopic observation mode, while 3 other telescopes are used in
stand-alone mode to monitor 3 other sources.

\begin{figure}[t]
  \begin{center}
    \includegraphics[width=34.5pc]{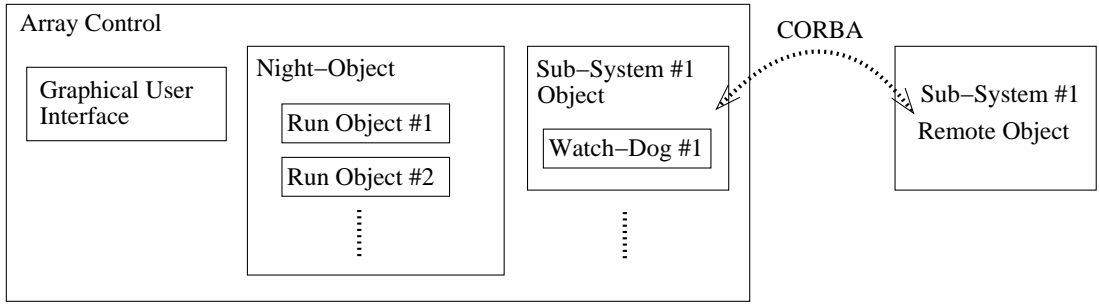}
  \end{center}
  \vspace{-0.5pc}
  \caption{Object oriented design of the Array Control program.
The user steers the telescopes with the help of a Graphical
User Interface (GUI) that also displays diagnostic information
about the performance of the telescopes and about the 
gamma-ray emission of the observed sources.
The telescopes can be devided into several sub-groups, 
each one observing a different astrophysical source.
So-called run-objects invoke observation runs with different
telescope sub-groups.
For each major software (and hardware) component, a sub-system
objects remotely invokes object methods via CORBA and 
monitors their status with watch-dog processes.
}
\label{des}
\end{figure}

The design of the Array Control and Telescope Control programs is shown
in Fig.\ \ref{des}. 
The software is written in C++, using the {\em Zthread} library, version 2.2.10
$\left[1\right]$ for the implementation of concurrent processes.
The communication between the distributed software components is facilitated 
by the {\em omniORB3.0} implementation $\left[2\right]$ of the CORBA protocol.

The user works with a Graphical User Interface 
(GUI) based on the {\em Qt}-package, version 3.1.1 $\left[3\right]$
that allows him or her 
to start server on the remote computers, to initialize and shut down
sub-systems, and to take calibration and science data.

The software is fully object oriented with a ``night-object''
performing diurnal initialization and shut down processes, 
``run-objects'' that log run-information in the 
{\em MySQL} data base $\left[4\right]$
and start and stop sub-systems, as well as ``sub-system objects''.
The latter include watch-dog processes that poll the status 
and key-performance parameters of a sub-system in regular time 
intervals to monitor their proper functioning.

Error situations are handles based on CORBA time-outs, function
return values and a well defined exception class.
\section{Status and Outlook}
The Array Control and Telescope Control programs described in this
paper will be used to take data with the first VERITAS prototype 
telescope in September, 2003. Subsequent development will aim at 
the full implementation of the multi-telescope capabilities, and
at the improvement of the display of diagnostic information that
will help the shift personnel to locate and remedy irregularities.
It is foreseen that between 1 and 2 persons will steer one
sub-system using one or more terminals.
Comprehensive display of the results from the 
online-analysis will enable the shift crew to rapidly
react to the detection of gamma-ray flares.
\section{References}
\vspace{\baselineskip}
\re
1.\ Eric Crahen, http://zthread.sourcefog.net
\re
2.\ http://omniorb.sourceforge.net/
\re
3.\ http://www.trolltech.com/products/qt/
\re
4.\ http://mysql.he.net/
\re
\endofpaper
\end{document}